# Volunteered Geographic Information and Computational Geography: New Perspectives[*]

Bin Jiang
University of Gävle, Gävle, Sweden
bin.jiang@hig.se

**Abstract** Volunteered geographic information (VGI), one of the most important types of user-generated web content, has been emerging as a new phenomenon. VGI is contributed by numerous volunteers and supported by web 2.0 technologies. This chapter discusses how VGI provides new perspectives for computational geography, a transformed geography based on the use of data-intensive computing and simulations to uncover the underlying mechanisms behind geographic forms and processes. We provide several exemplars of computational geography using OpenStreetMap data and GPS traces to investigate the scaling of geographic space and its implications for human mobility patterns. We illustrate that the field of geography is experiencing a dramatic change and that geoinformatics and computational geography deserve to be clearly distinguished, with the former being a study of engineering and the latter being a science.

## 1. Introduction

The field of geographic information science is currently benefiting from the increasing availability of massive amounts of volunteered geographic information (VGI) (Goodchild, 2007; Sui, 2008) contributed by individuals in the form of user-generated content supported by web 2.0 technologies. The emergence of VGI represents something of a paradigm shift in terms of geographic data acquisition from the conventional top-down approach, mainly dominated by national mapping agencies, to the bottom-up approach, in which data are contributed by individual volunteers through crowdsourcing - a massive collective of amateurs performing functions that were previously performed by trained professionals (Howe, 2009). Massive amounts of VGI of various types and the computations performed with these data constitute a significant part of eScience or data-intensive computing, which is being characterized as the fourth paradigm in scientific discovery (Hey et al., 2009). Among many others, OpenStreetMap (OSM) is one of the most successful examples of VGI.

OSM is a wiki-like collaboration, or a grassroots movement, to provide an editable map of the world using data from portable GPS devices, aerial photography and other free sources (Bennett, 2010). OSM is not owned by anyone, which is both amazing and unprecedented. Currently, there are more than 400,000 registered OSM contributors or users, and this number has been growing exponentially in the past few years. For the first time in human history, researchers can obtain street data of the entire world for analysis and computation. This analysis and computation can provide deep analytical insights into cities and our environments for sustainable development. This opportunity is significant and is very different from what is possible with Google Maps. Google Maps allows mashups, but its licensed and copyrighted data prevent us from obtaining analytical insights. We cannot learn how cities or regions have been sprawling outward by exploring only Google Maps. Instead, we need to perform analysis and computation to quantify the level of urban sprawl. In this regard, OSM, rather than Google Maps, freely provides a rich data source for researchers to use to better understand our cities and environments through advanced spatial analysis and computing. This understanding can further be used for spatial planning, e.g., redeveloping parts of a city or restricting further development of some parts of a country. In other words, OSM data can be analyzed to obtain knowledge in various forms of patterns, structures, relationships, and rules for spatial decision making. For instance, how is urban sprawl related to economic activities, population density and public health issues such as obesity?

---
[*] To appear as a chapter in: Sui D., Elwood S. and Goodchild M. (editors, 2012), *Crowdsourcing Geographic Knowledge: Volunteered Geographic Information in Theory and Practice*, Springer: Berlin

This chapter will discuss how geospatial analysis and computation of OSM data will lead to some hidden and surprising findings about the structures and patterns of geographic space. Our discussion is based on the assumption that OSM data, or VGI in general, are good enough to be used for computing and analysis. Although there are quality issues with VGI, this does not prevent us from uncovering hidden or surprising patterns about cities, environments and human activities. The accumulation of evidence, as well as Linus' law—*"given enough eyeballs, all bugs are shallow"* (Raymond, 2001, p. 30)—indicates that the quality of OSM data matches that of the data provided by mapping agencies, mirroring the results of studies (Giles, 2005) on other user-generated content such as Wikipedia.

This chapter will briefly review computational geography and its evolution along with other related notions emerging in the field of geographic information systems (GIS). We provide a new definition of computational geography, and we differentiate it from geoinformatics. We then present some exemplars of computational geography that rely on OSM data to uncover the underlying structures and patterns of geographic space. This chapter concludes with a few remarks.

## 2. What Is Computational Geography?

The notion of computational geography first appeared in 1994, when The Centre of Computational Geography was established as an interdisciplinary initiative at the University of Leeds. Two year later, in 1996, an international conference series on computational geography (geocomputation) was established. This conference has since been held more than 10 times. As seen in the literature, geocomputation has apparently become a favored term, although both geocomputation and computational geography refer to the same scientific understakings and have been interchangeably used.

What is computational geography? This question has been intensively studied and hotly debated in the GIS/geocomputation community (e.g., Longley et al., 1998; Gahagen, 1999; Ehlen et al., 2002). The same question has been answered and examined by various scholars on numerous occasions. To summarize, there are three basic views that were put forward by early pioneers on what computational geography is. The first view, mainly held by Stan Openshaw (2000) and Mark Gahagen (1999), recognizes the impact of increasing computing power and complex computational methods on geography or on the geosciences in general. This view stresses dealing with unsolvable geographic problems using, for example, high-performance computing, artificial intelligence, data mining and visualization. The second view is more concerned with the science of geography in a computationally intensive environment, and expects geocomputation to offer a means of explaining geographic phenomena. This view is mainly held by Helen Couclelis (1998) and Bill Macmillan (1998).

Paul Longley and his associates, such as Mike Goodchild (Longley et al., 2001), hold the third view that geocomputation is synonymous with geographic information science (GIScience), the science behind GIS technologies, which deals with fundamental questions raised by the use of systems and technologies.

The emergence of computational geography occurred at a time when GIS/geoinformatics as a tool underwent rapid development after a few decades of evolution and applications. Many GIS pioneers had started to think of certain fundamental issues surrounding the development of GIS. Along with computational geography, the widely recognized terms of geographic information science (Goodchild, 1992) and spatial information theory (Frank et al., 1992) appeared at almost the same time in the 1990s. It is no wonder that many GIS researchers see an overlap between GIS, geoinformatics, geomatics, geographic information science and spatial information theory. The co-emergence of these terms is a clear indictor that this field has been under rapid development and evolution. Every term is used to capture some essential part of the development. In what follows, we offer a slightly new definition of computational geography which captures the impact of data-intensive computing or eScience, and we state how it is different from, for example, geoinformatics.

Computational geography is a transformed data-driven geography that aims to understand the underlying mechanisms of geographic forms and processes via simulation of complex geographic phenomena and based on data-intensive computing. Along with the emerging field of computational social science (Lazer et al., 2009), computational geography is both data and computationally intensive. Computational geography is a science of geography that focuses on geographic forms and processes and that offers explanations through simulations. In other words, what computational geography seeks to answer is not only how the world *looks*, but also how the world *works*. This is in contrast to the focus of GIScience on how the world *looks* rather than how the world works (Goodchild, 2004). On the other hand, geoinformatics is an engineering geography, or geoscience in general, oriented toward the construction of tools and models for geospatial data acquisition, management, analysis and visualization to deal with real world problems. Despite the difference, both computational geography and geoinformatics are closely related in terms of geospatial information and developed tools. This view on the difference between geoinformatics and computational geography mirrors a similar view about "bioinformatics" being a field of engineering and "computational biology" being a science.

Much of the appeal of computational geography in the 21$^{st}$ century lies in the increasing availability of massive amounts of data about our environments and human activities in both physical and virtual spaces. With the increasing volume of data being generated from all types of scientific instruments, often acquired on a 24/7 basis, computational geography should adopt data-intensive geospatial computing to practice the science of geography. In this regard, the deployment of high-performance computing, grid/cloud computing and geographically distributed sensors provides a powerful means of computing. At the same time, the emerging VGI contributed by volunteers and gathered via social media constitutes a valuable and unprecedented data source for researchers in computational geography. In the next section, we will draw upon some of our recent studies to illustrate what computational geography is and how VGI can support computational geography research.

### 3. Examples of Computational Geography

Here, we describe a few recent computational geography studies that use VGI. Central to these studies are two basic concepts: topology and scaling. Topology refers to the topological relationships of numerous geographic units, while scaling is often characterized by a power law distribution or a heavy-tailed distribution in general. We illustrate in this section that both topology and scaling help uncover the underlying structures and patterns of geographic space, but first we must further clarify these two concepts.

**3.1 Concepts of Topology and Scaling**

Topology, initially a branch of mathematics, can be defined as the study of qualitative properties that are invariant under distortion of geometric space. For this reason, topology is also called "rubber geometry". In the GIS literature, the concept of topology has appeared on at least two occasions. The most familiar is probably topologically integrated geographic encoding and referencing (TIGER). The TIGER data structure or database was created by the US Census Bureau in the 1970s. The concept of topology also appeared in the GIS literature with Max Egenhofer and Robert Franzosa's formulation of the topological relationship (1991). Although the essence of topology is the same (which is about relationships), we adopt the notion of topology to refer to topologically based geospatial analysis. Topology, in contrast to geometric aspects such as locations, orientations, sizes, and shapes, is concerned with the relationship of geographic objects or units. To further elaborate the difference between topology and geometry, let us examine the London underground map as an illustrative example.

Figure 1 illustrates two versions of the London underground map: the left is geometrically corrected (a geometric map), while the right is topologically retained but geometrically distorted (a topological map). As one can see, all the locations and links of the stations are completely distorted in the topological map, with the exception of the relative orientations between the stations. This topological map is much more informative than the geometric map in terms of navigation along the tube lines. However, if we want to obtain in-depth structures or patterns, this topological map provides little information. For example, how many lines must be passed to get from one station to another? One can simply figure out the answer with the 12 tube lines. What if there are hundreds of lines? Put more generally, how many intermediate streets must be crossed in one city to go from one street to another? This is a basic question that, for example, is relevant to a taxi driver who is seeking optimal routes.

Figure 2 presents two versions of a topological map. Figure 2b illustrates a topological map showing the intersection or topology of tube lines, from which we can see certain in-depth structures. Among the 12 lines, 10 form an interconnected core in which nearly every one is connected to every one else, forming a sort of complete graph. Two lines, the East London line and the Waterloo & City line, are outside the core, with few connections to the others. This map indicates that if someone travels from the East London line to the Waterloo & City line, he or she must pass through another intermediate line; the two lines are not directly connected. In comparison with the geometric map, the topological map of the stations still retains certain geometric aspects, such as the relative positions and/or orientations of the lines and stations. In this regard, the topological map of the lines is purely topological: there are neither geometric locations for the nodes nor geometric distance for the links.

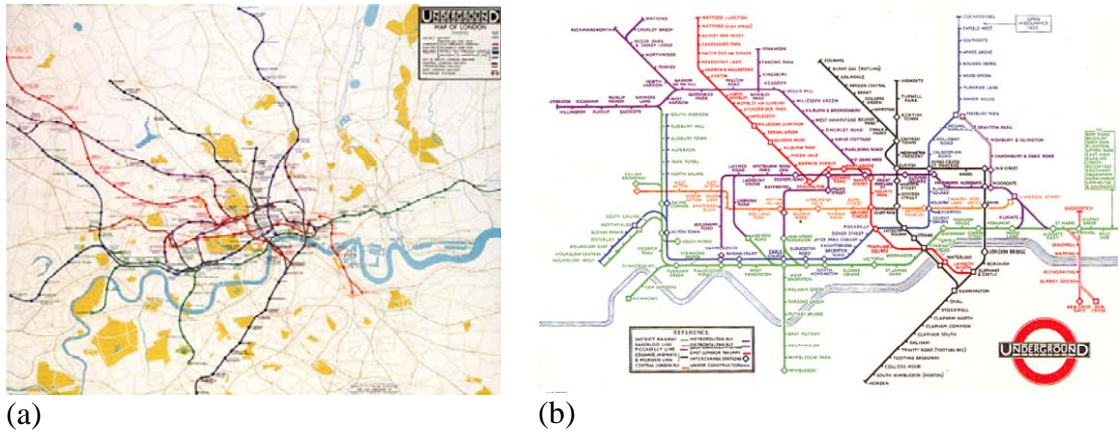

(a)                                                                         (b)

Figure 1: (Color online) Two versions of the London underground map: (a) a geometric map and (2) a topological map

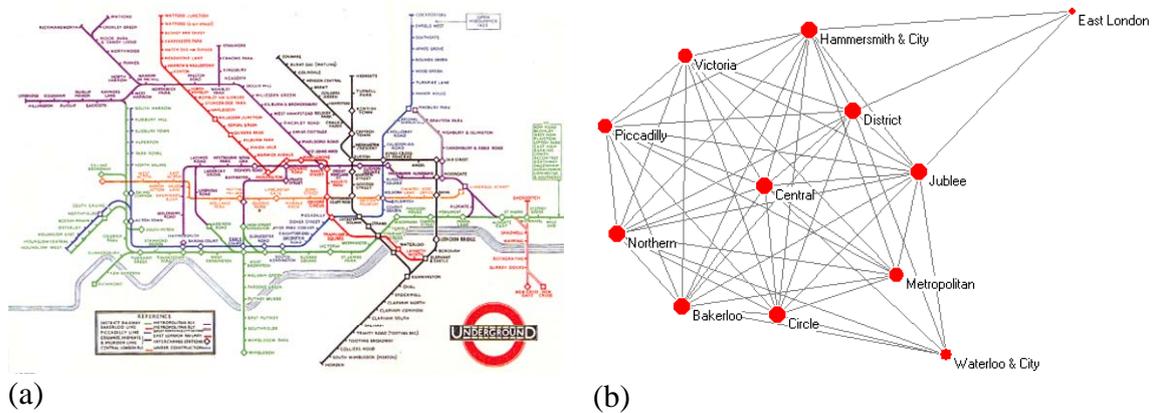

(a)                                                                         (b)

Figure 2: (Color online) Two versions of topological maps: (a) topology of stations and (b) topology of lines
(Note: the node sizes in the right image show how many other lines intersect, i.e., the degree of connectivity)

Scaling, or more specifically the scaling of geographic space in the context of this chapter, refers to the fact that there are far more small things than large ones in a geographic space. For example, there are far more small cities than large ones, far more short streets than long ones, and far more low buildings than high buildings. This phenomenon of "far more small events than large ones" is widespread, so it is said to be "more normal than normal". Scaling is the regularity behind many geographic phenomena. That there are far more small things than large ones also underscores a kind of spatial heterogeneity, i.e., there is no average thing in a geographic space. Because of the lack of an average thing, geographic space can also be said to be scale free. Note that scale in "scale free" means size, an average size or an arithmetic mean. "Scale free" implies that a notion of average size or mean make little sense in characterizing a variable that exhibits a power law distribution. The variation of things in a geographic space is highly heterogeneous or diverse. A major difference between the scaling of geographic space and spatial heterogeneity is that the former is characterized by a power law distribution, while the latter by a normal distribution. In general, scaling must be characterized by heavy-tailed distributions such as a power law or lognormal and exponential functions.

## 3.2 The First Example: Street Pattern of Sweden

The interconnected streets of a country constitute its basic infrastructure or backbone. Streets form a connected whole stretching across the county. Unfortunately, although the graph representation has found many applications in the computation of distance, routing and tracing, the underlying structure and pattern of streets cannot be simply illustrated with the conventional street networks using junctions and street segments, respectively, as the nodes and links of a graph. We call the conventional street network a geometric network in the sense that (1) every junction has a unique geographic location, and (2) every street segment is assigned a geometric distance. The geometric network embeds the connectivity of street junctions or that of street segments. Structurally, the conventional street network illustrates few interesting patterns because every junction is connected by almost the same number of other junctions, or equivalently, every segment is connected by almost the same number of other segments. However, the topology of streets exhibits a very interesting pattern, which can be said to be universal for all types of street networks all over the world.

We retrieved the entire Swedish street network database from the OSM databases and generated individual streets to assess how they are connected to each other. Note that the streets can be put into two categories: streets identified by unique names (Jiang & Claramunt, 2004) and natural streets generated by joining principles (Jiang et al., 2008). In this study, we first merged adjacent street segments with the same names to create street units and then adopted some principle to join the street units into natural streets. This procedure was performed because of the missing names for many street segments in the OSM databases. The resulting natural streets are very close to named streets. Eventually, we obtained over 160,000 streets from over 600,000 arcs. Figure 3 illustrates the hierarchical levels of the street network, and indicates that there are far more short streets (blue) than large ones (red). The least connected streets have a degree of 1, while the most connected streets have a degree of 1040. This very high ratio of the most connected degree to the least connected degree is a clear indicator of a heavy-tailed distribution.

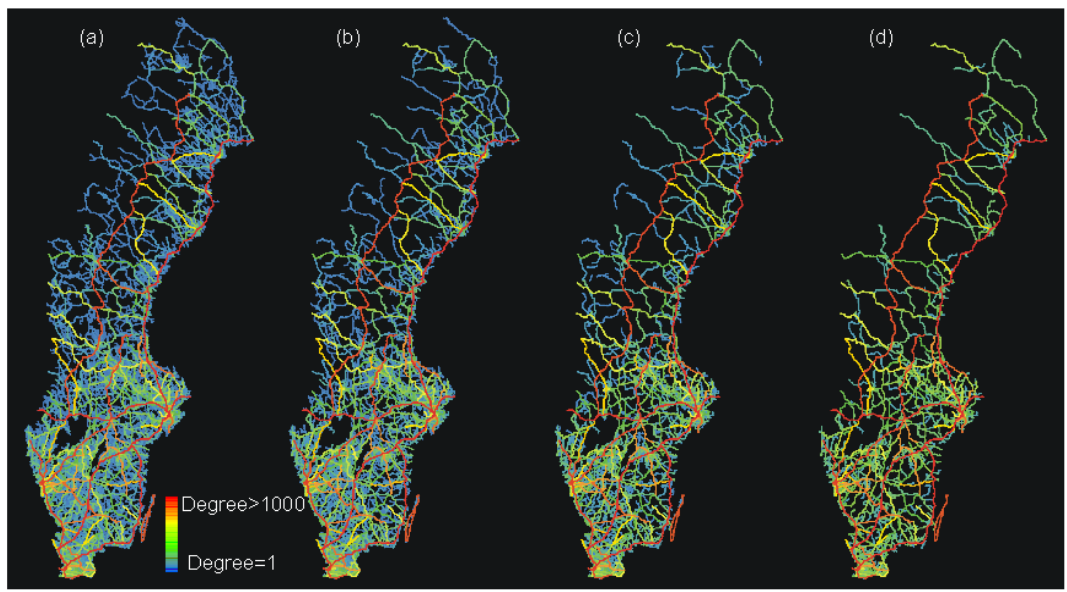

Figure 3: (Color online) Four levels of detail showing the hierarchical structure of the 160,000 streets of Sweden: (a) source map, (b) first level, (c) second level and (d) third level

This finding of the scaling pattern of the street network has profound implications for understanding other phenomena such as traffic flow. For example, the majority of traffic flow occurs on only a few of the most connected streets, while a vast majority of less connected streets accommodate only a small amount of the traffic flow (Jiang, 2009). Eventually, traffic flow and human mobility patterns also demonstrate this scaling pattern. We can further claim that it is the scaling of geographic space that shapes human movement patterns. This is the type of mechanism that we seek to discover through computational geography.

In addressing why human activities show the scaling pattern, Barabási (2010) tried to seek an answer from the perspective of people rather than that of space. He explains that we conduct our affairs in bursts because we set a priority for them. In terms of human movement, we spend most of our time (e.g., 90% of our time) in one place near our home, city or nearby, and only occasionally (e.g., 10% of our time or less) do we travel somewhere far from where we typically are. This is a traditional way of thinking that society is complicated because it consists of complicated persons; in fact, we can think of individuals as molecules or atoms (Buchanan, 2007). In a recent study, Jiang & Jia (2011a) created two types of moving agents (random and purposive) and simulated their movement patterns in a street network. It was found that moving behaviors have little effect on the overall traffic patterns.

Given the scaling pattern or property, cartographic generalization or mapping in general can be conducted in a simple manner. The head/tail division rule that we formulated can be applied in this case. The head/tail division rule states that anything with the scaling pattern can be divided into two imbalanced parts: a low percentage of larger items in the head and a high percentage of smaller items in the tail (Jiang & Liu 2011). In fact, Figure 3 above illustrates an application of the head/tail division rule by simply placing larger streets in the head recursively to create different levels of detail (Jiang, 2012). We further conjecture that the scaling of geographic space is something fundamental underlying cartographic generalization or mapping, and it is the underlying property that makes generalization and mapping possible.

**3.3 The Second Example: Street Block Pattern of France**

The second exemplar concerns the scaling pattern that emerges from the numerous street blocks of a country. A street block refers to a minimum ring or cycle formed by adjacent street segments, also called a "city block" in an urban environment. By street blocks, we mean both city blocks in cities and field blocks in the countryside. We developed a recursive algorithm to automatically derive a massive number of street blocks from street networks of the three largest European countries (Jiang & Liu, 2011). In this chapter, we use the French case as an example to illustrate how the interconnected blocks help uncover underlying patterns. First, we found that the sizes of the blocks show a lognormal distribution, one of several heavy-tailed distributions. This observed distribution implies that there are far more small blocks than large ones. Interestingly, using the head/tail division, we can partition all the blocks into categories: those smaller than the mean and those greater than the mean. In fact, the smaller blocks can be clustered into cities, while the larger ones constitute the countryside.

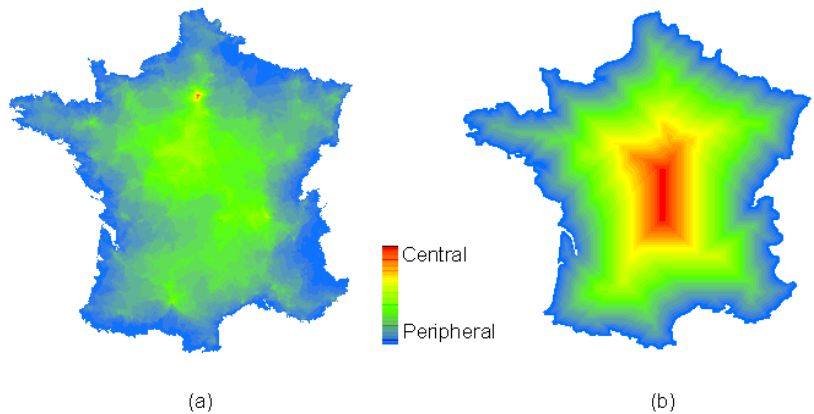

Figure 4: (Color online) Centers of France: (1) topological center and (2) geometric center

Second, we defined the notion of border numbers, indicating how far individual blocks are from the outermost border. Inspired by the notion of Bacon numbers, which show how far an actor or actress is from Kevin Bacon in the Hollywood universe ("six degrees of Kevin Bacon"), the border numbers are defined as follows. The blocks on the outermost border have border number 1, those blocks directly connected to blocks with border number 1 have border number 2, and so on. The border number is defined from a topological perspective, which is clearly different from a geometric perspective. Figure 4 illustrates the difference. Both geometric and topological distances are colored using a spectral color legend: the farther a block is, the more central it is. There are two centers: a topological center and a geometric center. Clearly, the topological center is the location of Paris. The geometric center is, in fact, a direct application of the medial axis (Blum, 1967). The geometric center is not what human beings perceive to be the center of the country, but the topological center is.

We can extend this reasoning to define a center in biological organisms. For example, what is the center of the human body? Relying on Blum's medial axis, we would derive the skeleton, but common sense tells us that both the heart and the mind are the two centers of the human body. We conjecture that if we were to take the topological perspective, we would be able to derive these two centers. This is based on the assumption that the sizes of the cells or any subunits, similar to the blocks, are heavy tail distributed. We have not found any scientific literature to support the above reasoning, but we need revised geographic imaginations in the computer age (Sui, 2004); data intensive computing, involving a massive amount of geographic information, facilitates creative imagination in some unique ways.

**3.4 The Third Example: Verifying Zipf's Law via Natural Cities**

The scaling property has several variants, one of which is Zipf's law, formulated by the linguist George Kingsley Zipf (1949). Zipf's law states that the size of any city is inversely proportional to its rank, e.g., the second largest city is 1/2 of the largest one; the third largest city is 1/3 of the largest, and so on. Usually, city sizes are measured by their population or physical extent. Conventionally, cities are defined legally or administratively, e.g., as census-designated places, urban areas, or metropolitan areas. The subjective and even arbitrary nature of these definitions poses problems for the verification of Zipf's law. In this regard, there have already been studies seeking more objective definitions of cities or city boundaries (e.g., Holmes & Lee, 2009; Rozenfeld et al., 2011). However, such studies still used aggregated data rather than individual-based data for defining cities.

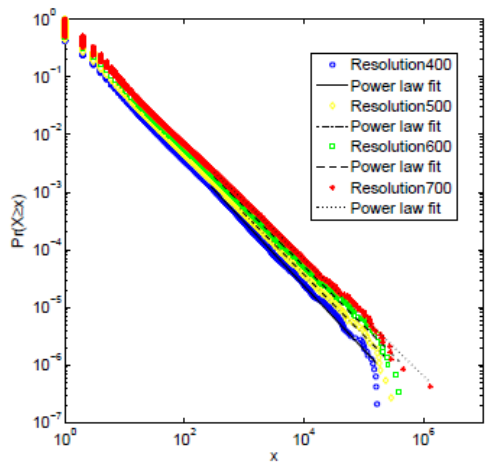

Figure 5: (Color online) Power law distributions of natural cities derived with different resolutions

We propose a new approach to defining cities by adopting street nodes as a proxy of population (Jiang & Jia, 2011b). We retrieved over 120 GB of OSM data for the USA, and we extracted 25 million street nodes. By applying a clustering algorithm, we grouped the nearest street nodes into individual urban settlements. We define the derived urban settlements as "natural cities" because the clustering was performed recursively and automatically. Eventually, we obtained approximately 2-4 million natural cities, depending on the chosen resolutions for the clustering process. The resolutions we chose were 400, 500, 600, and 700 meters because they were the same magnitude as the city block sizes. Interestingly, the derived natural cities strikingly exhibit a power law distribution, but the Zipf exponent may deviate from 1.0.

We conducted a comparison study by examining Zipf's law between natural cities and urban areas and found that the law is remarkably stable for all natural cities, ranging from the largest to the smallest (with only one road node) (Jiang & Jia, 2011b). Surprisingly, the Zipf exponent remains unchanged for the entire range; one can refer to Figure 5 for the log-log plot of the distribution. This result contrasts sharply with the results for urban areas, which exhibit Zipf's law for some of the largest cities; the Zipf exponent varies from one part to another of the entire range. This behavior may indicate that Zipf's law, or the power law in general, underlies certain self-organized processes.

## 4. Discussions

We have introduced two key concepts around which three studies regarding computational geography have been presented to uncover the underlying scaling property, and in particular to illustrate the underlying mechanism of human activity patterns in geographic space. Let us further elaborate the implications of these studies. Current geospatial analysis is very much dominated by two stubborn mindsets: one is geometric thinking in terms of sizes, shapes, orientations and positions and the other is a Gaussian way of thinking that uses geostatistics to characterize spatial properties such as spatial dependence and spatial heterogeneity. Spatial dependence also goes by the name spatial autocorrelation, which is expressed succinctly by the first law of geography—*"everything is related to everything else, but near things are more related than distant things"* (Tobler 1970, p. 236). Spatial heterogeneity emphasizes that the Earth's surface is heterogeneous, but this heterogeneity is still very much characterized by a normal distribution. Statistical theories based on scaling and heavy-tailed distributions are rarely adopted for the study of geographic phenomena. We therefore want to promote two alternative ways of thinking, topological thinking and scaling thinking, to get insights into geographic forms and processes.

Topological thinking is rooted in one of the two fundamental views about space – Leibniz's relative space, which focuses on relationships between individual objects. This is to be understood in contrast to geometric or topographic thinking, which is dominated by Newtonian ideas about absolute space. The topological focus on relationships is not particularly new, since it was well treated in geographic literature a long ago (e.g., Haggett & Chorley, 1969). However, what is considered to be unique or interesting in this chapter is how topological thinking helps us appreciate the scaling pattern. Thinking topologically is part of thinking spatially, and it is becoming increasingly important at a time when most of human activities have shifted to virtual space and when emerging social media is coming to dominate everyday human activities (Allen, 2011).

It should be noted that there are at least two factors that prevent us from gaining insight into the scaling of geographic space: (1) how we look at geographic space (perspective) and (2) the size of the study areas we choose (scope). For example, the geometric representation of street networks is unable to demonstrate the scaling property. More importantly, the geometric representation is not what human beings perceive about street networks. The second factor is closely related to the availability of massive amounts of crowd-sourced geographic information. In this regard, VGI provides an unprecedented data source enabling us to conduct this type of geospatial analysis and modeling.

One may argue that geographic information maintained by governments or private companies would allow us to achieve the same goal. This would indeed be true if the data were accessible but this is easily said and only rarely done. First of all, there are many restrictions on access to data. Secondly, the data studied are rarely shared among all interested parties. What if some other researchers want to verify the results with the same data? Often there is no way to do so. This is a big constraint for scientific research. With OSM, for the first time, we can seamlessly integrate publicly available geographic information, and continuously keep it updated through volunteered efforts.

There are many other formats of VGI, emerging from a variety of social media such as Facebook, Twitter, and Flickr, which are not addressed in this chapter. These data may go beyond what Mike Goodchild (2007) refer to about VGI, but their potential for studying spaces and places could be enormous. The convergence of GIS and social media is providing new ways of studying interactions among people, space and place, which are fundamental to geography (Sui & Goodchild, 2011). We believe there are many studies to be done in the future.

## 5. Concluding Remarks

In this chapter, we briefly reviewed the emergence of computational geography, its definitions and its evolution in the past two decades. We provided an alternative definition of computational geography that deals with simulation of geographic phenomena to uncover the underlying mechanisms of geographic forms and processes. Computational geography is distinct from geoinformatics, which is more concerned with the engineering side of geographic information in terms of data acquisition, management, analysis and visualization. Governments have put enormous effort into data collection (e.g., population censuses, housing and economic activity), but these data are seldom made available for research. In this regard, VGI, and OSM in particular, provides a valuable data source for computational geography.

We provided several examples of research in computational geography that relies on VGI and that rests on conceptualizations of relative and relational space. Despite of being an important way to conceptualize space, geographic studies still tend to examine (absolute) space in the geometric sense rather than the topological sense we refer to in the paper. It is time, in the context of computational geography, to rethink the relative view of space in geographic studies (Gatrell 1984). We have seen from the chapter that topology and scaling indeed matter in geospatial analysis. We need to shift our mindsets from geometric to topological thinking and from the Gaussian mindset to something that is more normal than normal - the scaling property.

**Acknowledgement**

The author would like to thank the book editors for their constructive comments that led to many improvements, in particular Daniel Sui, for bringing my attention to recent references on topological thinking in human geography.

1737-1748.

Tobler, W. (1970). A computer movie simulating urban growth in the Detroit region. *Economic Geography*, 46(2), 234-240.

Zipf, G. K. (1949). *Human behavior and the principles of least effort.* Addison Wesley: Cambridge, MA.